# Bose-Einstein Correlations of Pion Wavepackets

H. Merlitz[1,2] and D. Pelte[2,3]

[1] Gesellschaft für Schwerionenforschung Darmstadt
[2] Physikalisches Institut der Universität Heidelberg, D-69120 Heidelberg
[3] Max-Planck-Institut für Kernphysik, D-69117 Heidelberg

**Abstract.** A wavepacket model for a system of free pions, which takes into account the full permutation symmetry of the wavefunction and which is suitable for any phase space parametrization is developed. The properties of the resulting mixed ensembles and the two-particle correlation function are discussed. A physical interpretation of the chaoticity $\lambda$ as localization of the pions in the source is presented.

Two techniques to generate test-particles, which satisfy the probability densities of the wavepacket state, are studied:

1. A Monte Carlo procedure in momentum space based on the standard Metropolis technique.
2. A molecular dynamic procedure using Bohm's quantum theory of motion.

In order to reduce the numerical complexity, the separation of the wavefunction into momentum space clusters is discussed. In this context the influence of an unauthorized factorization of the state, i.e. the omission of interference terms, is investigated. It is shown that the correlation radius remains almost uneffected, but the chaoticity parameter decreases substantially. A similar effect is observed in systems with high multiplicities, where the omission of higher order corrections in the analysis of two-particle correlations causes a reduction of the chaoticity and the radius.

The approximative treatment of the Coulomb interaction between pions and the source is investigated. The results suggest that Coulomb effects on the correlation radii are not symmetric for pion pairs of different charges. For $(\pi^-,\pi^-)$ pairs the radius, integrated over the whole momentum spectrum, increases substantially, while for $(\pi^+,\pi^+)$ pairs the radius remains almost unchanged.

## 1 Introduction

It is a well confirmed experimental fact [1, 2] that pions, which are produced in heavy ion collisions, display Bose-Einstein correlations. These correlations are a consequence of the quantum mechanical interference in the corresponding symmetrical $n$-particle wave function. They contain a large amount of information about the statistical properties of the momentum- *and* configuration space distribution of the system, and thus provide a method to probe the source geometry. In contrast to the situation found in astronomy, where Hanbury Brown and Twiss first measured the radii of stars by analyzing the correlation functions of photons [3], the pion sources in heavy ion collisions are of higher complexity. Computer simulations are necessary to interpret the pion correlation functions. In addition such simulations have to take into account the constraints imposed by the experiment. Adequate models therefore have to be versatile and flexible enough not only to allow for a proper treatment of the quantum effects, but also have to be based on a realistic description of the reaction and to include the experimental conditions.

The strategy to investigate the correlation phenomena in heavy ion physics is displayed in the flow diagram Fig. 1:

1) The goal of the source generator (Box 1) is to produce the phase space distribution of the pions at *freeze-out* time, which is defined as the moment at which the strong interactions have ceased to exist. Since the fireball, which is the dominant source of pions, is not static but expands after being created in a state of highest density and temperature, and since the pions are probably emitted at every instant during the expansion phase, the freeze-out time represents an individual property of each pion. Although there exist codes which use a parametrization of the pion - phase space distribution (Box 1a), a more realistic description of the pion - emission process is realized in codes which are based on microscopical models like BUU or QMD [4, 5] and which do not provide parametrical expressions for the source (Box 1b).

2) After they are produced by the source generator, the pions still do not exhibit Bose-Einstein correlations, since the presently available codes do not include quantum statistics. The first step to obtain correlated pions is to construct a symmetrized wavefunction (Box 2). This

*Correspondence to:* holgi@pel4.mpi-hd.mpg.de

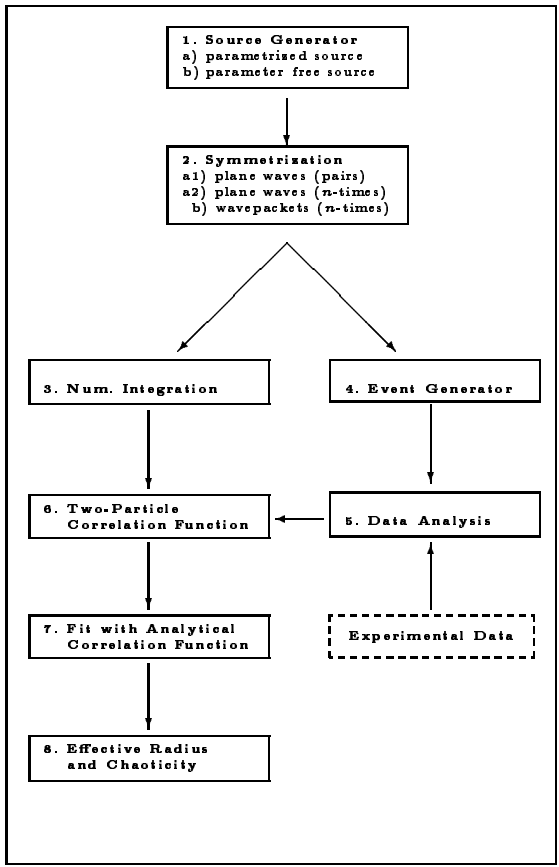

**Fig. 1.** The procedures needed to obtain effective correlation parameter (Box 8) from source models (Box 1).

can be done in different ways: The most common method is to use the two-particle plane wave state in momentum space (Box 2a1)

$$\Phi(\mathbf{p}_1, \mathbf{p}_2) \sim \; e^{-i\,\mathbf{p}_1 \cdot \mathbf{x}_1/\hbar} \; e^{-i\,\mathbf{p}_2 \cdot \mathbf{x}_2/\hbar}$$
$$+ \; e^{-i\,\mathbf{p}_1 \cdot \mathbf{x}_2/\hbar} \; e^{-i\,\mathbf{p}_2 \cdot \mathbf{x}_1/\hbar} \; , \qquad (1)$$

where $\mathbf{p}_i$ denotes the momentum and $\mathbf{x}_i$ the coordinate of the $i$'th pion. As soon as the phase space density increases, higher order correlations begin to contribute to the statistics of the pions and a symmetrization of $n$'th order becomes the more adequate description of the quantum state (Box 2a2).

3) Once the source properties are defined and the wavefunction is symmetrized, the 2-particle correlation function in momentum space

$$\mathcal{C}_{12}(\mathbf{p}_1, \mathbf{p}_2) = \frac{\mathcal{P}_{12}(\mathbf{p}_1, \mathbf{p}_2)}{\mathcal{P}(\mathbf{p}_1)\,\mathcal{P}(\mathbf{p}_2)} \qquad (2)$$

can be constructed numerically. Here, $\mathcal{P}(\mathbf{p})$ is the single-particle momentum distribution, $\mathcal{P}_{12}(\mathbf{p}_1, \mathbf{p}_2)$ is the probability to find simultaneously two particles with momenta $\mathbf{p}_1$ and $\mathbf{p}_2$. There exist a couple of numerical codes which are based on the plane wave approximation (Box 2a1) and additionally include approximative treatments of final state interactions via partial wave expansions [6].

4) Alternatively it is possible to produce correlated pion distributions by means of a quantum mechanical event generator. At present there exist two techniques for the eventwise generation of correlated pions: In the codes of the first type the pion pairs are weighted with the factor [7]

$$1 + \cos\left((\mathbf{x}_i - \mathbf{x}_j) \cdot (\mathbf{p}_i - \mathbf{p}_j)/\hbar\right) \; , \qquad (3)$$

which is the statistical weight for two particles described by plane waves as in Eq. (1). Additional weights occur in case of interactions. In Sect. 2.3 we will discuss to which degree these methods suffer from the inconsistencies inherent in the plane wave picture, which is a semiclassical approximation of the quantum mechanical state. Another problem is due to the fact that correlations of higher order than 2 are ignored. This is not true for the event generators of the second type, which where developed to produce events of high multiplicities and which employ quantum statistical methods [8] or approximations for the $n$-particle state [9] (Box 2a2). These methods, however, are only applicable for the simplest source parametrizations (Box 1a) and can therefore not be combined with event generators based on microscopical models (Box 1b). All presently available methods have in common that they exclude an explicit time development of the pionic system, since the description via plane waves or quantum statistics implies the existence of stationary states.

5) The advantage of generated (hard) events as compared to a direct computation of the correlation function (Box 3) is that all analysis procedures, which originally were developed for the experimental data (dashed box), can also be applied to the simulated data. This includes detector simulations and tracking procedures and therefore allows for the best comparability with the experiment.

6) In general the comparison of the resulting correlation function Eq. (2) with the experimental function only tests the applicability of the model but does not yield values for the source parameters.

7) Therefore, an analytical model is employed to extract some characteristic source parameters. A simple parametrization of the pion source (Box 1) is the static fireball with a Gaussian density distribution, which we may call the *standard pion source* [9]:

$$\rho(\mathbf{r}) = \frac{1}{(\pi R^2)^{\frac{3}{2}}} \exp\left(\frac{-\mathbf{r}^2}{R^2}\right) \; . \qquad (4)$$

Note that $R$ is not the rms radius, instead $R/\sqrt{2}$ is the Gaussian $\sigma$ in each direction, so that the rms radius of the source is given by $\sqrt{3/2}\,R$. For such a source and when the particles are described as plane waves (Box 2a1) and chaotic emission is assumed, the corresponding correlation function can be evaluated analytically and yields [10]

$$C_{12}(q) = 1 + \exp\left(-\frac{R^2\,q^2}{2\hbar^2}\right) \; , \qquad (5)$$

where $q = |\mathbf{p}_2 - \mathbf{p}_1|$ is the magnitude of the momentum difference and the radius $R$ is used as fit parameter. "Chaotic" means that the emission coordinates of the particles are independent, i. e. the two-particle correlation function in *configuration* space is a delta distribution $\mathcal{C}_{12}(\mathbf{r}_1 - \mathbf{r}_2) = \delta(\mathbf{r}_1 - \mathbf{r}_2)$. This is a semiclassical approximation in the sense that quantum interference effects in configuration space are neglected. The picture of a chaotic source is a good approximation in astronomy, since the star radii are much larger than the coherence length of the emitted light, but it becomes inadequate in heavy ion physics, where the de Broglie wavelength of the pions is of the same order as the source dimension. In order to compensate for this oversimplification and to obtain a better agreement with the experimental results, a second fit parameter $\lambda$ is introduced, which is called *chaoticity parameter* and which accounts for a possible contribution of correlations in configuration space. The parameter $\lambda$ equals 1 in case of total chaoticity and equals 0 in case of total coherence. We may call

$$C_{12}(q) = 1 + \lambda \, \exp\left(-\frac{R^2 \, q^2}{2\hbar^2}\right) \qquad (6)$$

the *standard correlation function*.

8) Finally, the effective radius $R_{\text{eff}}$ and the chaoticity $\lambda$ are obtained by a fit of the analytical model Eq. (6) to the correlation function Eq. (2).

The interpretation of the radius parameters extracted from two-pion correlations under conditions present in heavy-ion collisions at ultrarelativistic energies (Box 7 in the flow diagram) has been the subject of many recent publications, see for example [12, 13, 14, 17, 16].

This work is intended to close some of the gaps still present in today's numerical modeling of pion correlations (Box 2) and event generation (Box 4). In Sect. 2 we present a quantum mechanical description of noninteracting many-pion systems in terms of wavepackets. A wavepacket model was already proposed by Sinyukov et al. [15], which, however, cannot be applied to pions with arbitrary momentum distribution, since it makes no distinction between classical- and quantum mechanical ensembles in momentum space. A more general treatment was presented by Padula et al. [16], who developed a generalization of Pratt's formula [17] in order to describe wavepackets in Wigner space. These models, however, neglect the dispersion of the wavepackets, so that a correct time dependent treatment in configuration space is not possible.

We will present a most general description which is independent on any parametrization in configuration- and momentum space (Box 1b) and includes $n$-particle correlations (Box 2b, Sect. 2). This model introduces a new characteristic parameter of the system, the *initial wavepacket width* $\sigma_o$. In Sect. 2.2.1 we discuss how this parameter effects the probability distributions of the pions in configuration- and momentum space. In Sect. 2.2.2 we discuss the correlation function of the wavepacket model and demonstrate how the finite width of the single-particle state accounts for partial coherence in a quite natural manner. A justification of the model in terms of physical arguments is offered in Sect. 2.3. In Sect. 3 a Monte Carlo procedure which allows us to produce events of correlated pions in full symmetry and independent of the source parametrization is presented. An approximative treatment of states with multiplicities higher than $\approx 10$ is discussed in Sect. 3.2, in Sect. 3.3 the influence of neglected interference terms on the 2-particle correlation function is investigated. In Sect. 4 a molecular dynamic procedure which allows for an explicit time development of the system in configuration- and momentum space is presented. First we demonstrate how Bohm's quantum theory of motion is suitable to extend the rules for the time development of a quantum mechanical ensemble average into rules for single representatives of the ensemble. Then, in Sect. 4.2 an approximative treatment of Coulomb interaction between pions and their source is presented.

Since our approach is of pure quantum mechanical nature, relativistic effects are not included, and a consistent way to extent the formalism so that it covers relativistic quantum mechanics is at present not available.

## 2 Generalized model

We consider a set of $n$ pions which may be produced by one of the usual source generators, e. g. BUU [4] or QMD [5], after freeze-out, so that no strong interactions occur at later times. The state of the system is defined by the phase space occupation

$$\{(\mathbf{X}_1, \mathbf{P}_1), (\mathbf{X}_2, \mathbf{P}_2), \dots, (\mathbf{X}_n, \mathbf{P}_n)\} \, . \qquad (7)$$

Naturally, these pions do not display Bose-Einstein correlations, since no symmetrization was taken into account in the course of the simulation. A large number of such states therefore build up a *classical ensemble*. The idea is first to symmetrize the system, construct a $n$-particle wavefunction $\Psi$ and then to generate a set of $n$ test-particles whose phase space distribution coincides with the probability density of $\Psi$. The coordinates of these test-particles will be written in small letters, so that

$$\{(\mathbf{x}_1, \mathbf{p}_1), (\mathbf{x}_2, \mathbf{p}_2), \dots, (\mathbf{x}_n, \mathbf{p}_n)\} \qquad (8)$$

denotes a single representative of a *quantum mechanical ensemble* which is defined by the pure state $\Psi$.

### 2.1 Wavepacket representation

Each of the $n$ pions defines a Gaussian wavepacket with the following configuration space representation:

$$\psi_i(\mathbf{x}, 0) = (2\pi\sigma_o^2)^{-3/4} \exp\left(\frac{i\mathbf{P}_i \cdot \mathbf{x}}{\hbar} - \frac{(\mathbf{X}_i - \mathbf{x})^2}{4\sigma_o^2}\right) \, . \qquad (9)$$

It is centred in $\mathbf{X}_i$ and the corresponding probability density

$$|\psi_i|^2 = (2\pi\sigma_o^2)^{-3/2} \exp\left(-\frac{(\mathbf{X}_i - \mathbf{x})^2}{2\sigma_o^2}\right) \qquad (10)$$

has a width $\sigma_o$ which is so far arbitrary. Its meaning will be examined in the following subsections. The solution of the Schrödinger equation yields:

$$\psi_i(\mathbf{x}, t) = (2\pi s^2(t))^{-3/4} \times$$
$$\exp\left(\frac{i\left(\mathbf{P}_i \cdot \mathbf{x}(t) - \frac{P_i^2 t}{2m}\right)}{\hbar} - \frac{\left(\mathbf{x}(t) - \mathbf{X}_i - \frac{\mathbf{P}_i t}{m}\right)^2}{4s(t)\sigma_o}\right) \qquad (11)$$

Certainly, the wavepacket moves with the group velocity $\mathbf{P}_i/m$, and, as a result of dispersion, has a time dependent width:

$$s(t) = \sigma_o \left(1 + i\frac{\hbar t}{2m\sigma_o^2}\right) . \qquad (12)$$

The $n$-particle wavefunction is the symmetrized product

$$\Psi(\mathbf{x}_1, \mathbf{x}_2, \ldots, \mathbf{x}_n, t) = \frac{1}{\sqrt{n!}} \sum_\sigma \psi_{\sigma(1)}(\mathbf{x}_1, t)\, \psi_{\sigma(2)}(\mathbf{x}_2, t) \\ \cdots \psi_{\sigma(n)}(\mathbf{x}_n, t) , \qquad (13)$$

sometimes called *permanent* in contrast to its antisymmetrical counterpart, the *determinant*, which is required to describe systems of identical fermions. The sum includes all $n!$ permutations $\{\sigma(1), \sigma(2), \ldots, \sigma(n)\}$. On account of simplicity we assume that all pions are emitted at the same time $t = 0$. Otherwise, we have to substitute $s(t) \to s(\tau_i)$ and $\mathbf{P}_i t \to \mathbf{P}_i \tau_i$, where $\tau_i$ is the time after production of the $i$'th pion.

The momentum space representation of a single particle is obtained by the Fourier-transformation

$$\phi_i(\mathbf{p}, t) = h^{-3/2} \int d^3x\, \psi_i(\mathbf{x}, t) \exp\left(\frac{-i\, \mathbf{p} \cdot \mathbf{x}}{\hbar}\right) \qquad (14)$$

$$= \mathcal{N}_1 \exp\left(-\frac{\sigma_o^2}{\hbar^2}\, d\mathbf{p}_i^2 + \frac{i}{\hbar}\left(\mathbf{X}_i \cdot d\mathbf{p}_i - \frac{\mathbf{p}^2 t}{2m}\right)\right) \qquad (15)$$

with

$$\mathcal{N}_1 = \left(\sqrt{\frac{2}{\pi}}\, \frac{\sigma_o}{\hbar}\right)^{3/2} \qquad (16)$$

and $d\mathbf{p}_i = \mathbf{P}_i - \mathbf{p}$. It displays the remarkable property of stationarity, i.e. there is no motion (since there is no final state interaction) or dispersion, only a rotating phase. The constant width is

$$\sigma_{\mathbf{p}} = \frac{\hbar}{2\sigma_o} . \qquad (17)$$

The $n$-particle wavefunction has to be constructed in the same way as in configuration space:

$$\Phi(\mathbf{p}_1, \mathbf{p}_2, \ldots, \mathbf{p}_n, t) = \frac{1}{\sqrt{n!}} \sum_\sigma \phi_{\sigma(1)}(\mathbf{p}_1, t)\, \phi_{\sigma(2)}(\mathbf{p}_2, t) \\ \cdots \phi_{\sigma(n)}(\mathbf{p}_n, t) . \qquad (18)$$

In a first step, we have described the $n$-particle distribution (7) by a symmetrical $n$-particle wavefunction $\Psi$ or $\Phi$, respectively. *From now on, the coordinates $(\mathbf{X}_i, \mathbf{P}_i)$ do not play the role of particle coordinates.* Instead, our task is to find a set of $n$ *test-particles*, whose phase space distribution (8) is defined by the probability densities $|\Psi|^2$ (for the coordinates) and $|\Phi|^2$ (for the momenta), respectively. Before methods to obtain these particles are presented, we first want to discuss the statistical properties of these test-particles, which can substantially deviate from the distributions of the classical ensemble (7).

## 2.2 Ensemble properties of the test-particles

We consider a simple physical situation: The configuration space distribution of the pion wavepackets is given by the standard source Eq. (4) and the momentum space distribution by a (classical) Maxwellian

$$f(\mathbf{P}) = (2\pi m T)^{-3/2} \exp\left(\frac{-\mathbf{P}^2}{2mT}\right) \qquad (19)$$

with a given temperature $T$. Of course, our description does not require to use a particular parametrization, but it will help us to develop an intuitive understanding of the model and provides us with the possibility to cross-check the results of the numerical codes.

### 2.2.1 Single-particle distributions

The momentum distribution of the test-particles is defined by the *mixed state*, which is obtained by integration of the single-particle (pure state) density $|\phi(\mathbf{p})|^2$ over the (classical) density distribution Eq. (19):

$$\mathcal{P}(\mathbf{p}) = (2\pi m T)^{-3/2} \int |\phi(\mathbf{p})|^2 \exp\left(\frac{-\mathbf{P}^2}{2mT}\right) d^3P$$

$$= (2\pi m T_{\text{eff}})^{-3/2} \exp\left(\frac{-\mathbf{p}^2}{2mT_{\text{eff}}}\right) . \qquad (20)$$

The distribution remains to be of Maxwellian type, but the temperature has increased to

$$T_{\text{eff}} = T + \frac{\hbar^2}{4m\sigma_o^2} \equiv T + T_q . \qquad (21)$$

The *quantum temperature* $T_q$ is a consequence of the zero point energy the particles gain when they are localized in the configuration space by wavepackets. Alternatively, one may regard the increase of the temperature as the result of a delocalization in momentum space.

The same calculation can be done for the configuration space distribution, yielding

$$\mathcal{P}(\mathbf{x}) = (\pi R^2)^{-3/2} \int |\psi(\mathbf{x})|^2 \exp\left(\frac{-\mathbf{X}^2}{R^2}\right) d^3X$$

$$= (\pi \tilde{R}^2)^{-3/2} \exp\left(\frac{-\mathbf{x}^2}{\tilde{R}^2}\right) \qquad (22)$$

with

$$\tilde{R} = \sqrt{R^2 + 2\sigma_o^2} . \qquad (23)$$

Certainly, the test-particles occupy a larger source volume than the wavepacket centers, a consequence of the finite wavepacket size. We conclude that the introduction of wavepackets leads to a blurring of the single-particle distributions: Localization of the test-particles in small wavepackets introduces a zero point energy which leads to higher temperatures, a consequence of Heisenberg's uncertainty relation. On the other hand, large wavepackets yield a better localization in momentum space and therefore do not disturb the momentum distribution, but then the information in configuration space is lost. Thus the optimal width $\sigma_o$ has to be a compromise between these two extremes.

### 2.2.2 Two-particle correlation function

The 2-particle correlation function is defined by Eq. (2), where the nominator is obtained by integrating the 2-particle pure state density over the (classical) phase space density distributions

$$\mathcal{P}_{12} = \int \frac{1}{\mathcal{N}_{12}} \rho(\mathbf{X}_1) \rho(\mathbf{X}_2) f(\mathbf{P}_1) f(\mathbf{P}_2) |\Phi(\mathbf{p}_1, \mathbf{p}_2)|^2 \\ \times d^3X_1 d^3X_2 d^3P_1 d^3P_2 . \quad (24)$$

We have

$$|\Phi(\mathbf{p}_1, \mathbf{p}_2)|^2 = \frac{1}{2} \left( |\phi_1(\mathbf{p}_1)|^2 |\phi_2(\mathbf{p}_2)|^2 \right. \\ \left. + |\phi_2(\mathbf{p}_1)|^2 |\phi_1(\mathbf{p}_2)|^2 + 2 Re\{\mathcal{I}\} \right) \quad (25)$$

with the interference term

$$\mathcal{I} = \phi_1(\mathbf{p}_1) \phi_2(\mathbf{p}_2) \phi_2^*(\mathbf{p}_1) \phi_1^*(\mathbf{p}_2) . \quad (26)$$

Notice that the interference term, in contrast to the single-particle probability densities, depends on momentum as well as on configuration space coordinates, which requires to integrate $|\Phi(\mathbf{p}_1, \mathbf{p}_2)|^2$ over the complete phase space. The norm $\mathcal{N}_{12}$ is

$$\mathcal{N}_{12} = \int |\Phi(\mathbf{p}_1, \mathbf{p}_2)|^2 d^3p_1 d^3p_2 = \\ 1 + \exp\left( \frac{-(\mathbf{P}_1 - \mathbf{P}_2)^2}{4\sigma_p^2} \right) \exp\left( \frac{-(\mathbf{X}_1 - \mathbf{X}_2)^2}{4\sigma_o^2} \right) . \quad (27)$$

$\mathcal{N}_{12}$ is also a function of the centre coordinates and has to be included in the integration Eq. (24). Physically, the second term in $\mathcal{N}_{12}$ denotes the degree of overlap in phase space. It aquires values between 0 and 1, thus the integral Eq. (24) can be solved in terms of a Taylor expansion of the inverse norm $\mathcal{N}_{12}^{-1}$ and integrated term by term. The resulting expressions are quite cumbersome and provide little physical insight. Therefore, in subsequent sections we will treat the integration numerically, but beforehand we may get some rough impression about the solution by means of approximations.

First we may examine the 0'th order Taylor contribution, i. e. the result of Eq. (24) in the approximation $\mathcal{N}_{12} = 1$. We obtain

$$\mathcal{C}_{12}(\mathbf{p}_1, \mathbf{p}_2) = 1 + \exp\left( \frac{-R_{\text{eff}}^2 (\mathbf{p}_2 - \mathbf{p}_1)^2}{2\hbar^2} \right) \quad (28)$$

with the effective radius

$$R_{\text{eff}} = \sqrt{R^2 + 2\sigma_o^2 \frac{T}{T_{\text{eff}}}} . \quad (29)$$

Two points are remarkable:

1. Like the standard correlation function Eq. (6), $\mathcal{C}_{12}$ only depends on the absolute momentum *difference* $|\mathbf{p}_2 - \mathbf{p}_1|$, but the chaoticity parameter $\lambda$ is fixed to 1. As we will see below, this changes if Taylor terms of higher order are taken into account.
2. The effective radius obtained by interferometry is not equal to the radius $\tilde{R}$ of the single-particle distribution Eq. (22). Instead, there is an additional contribution from the quantum temperature. We will see in Sect. 3 that higher order Taylor terms increase the value of $R_{\text{eff}}$.

Finally we demonstrate how the finite size of the wavepackets effects the chaoticity $\lambda$ for our model. A more general treatment of partial coherence is given in [15]. For our argumentation we adopt a quantum statistical point of view and assume that due to the overlap of wavepackets there exist a finite number of $K$ configuration space cells. Different cells represent independent states. Pions which are emitted from identical cells represent the coherent part of the correlation function and do not contribute to the chaoticity $\lambda$, while pions from different cells are independent and increase the chaoticity. Therefore we define

$$\lambda = 1 - \frac{\text{Number of coherent pairs}}{\text{Number of all pairs}} . \quad (30)$$

Employing the fact that the statistical weight of a cell, which is already occupied by $N$ bosons, is equal to $N+1$, $\lambda$ can be evaluated by means of combinatorial methods as shown in the appendix. We obtain

$$\lambda = 1 - \frac{2}{K+1} , \quad (31)$$

independent of the multiplicity of the system. In order to estimate $K$, we note that $\lambda$, following Eq. (6), can be expressed as

$$\lambda = \mathcal{C}_{12}(q = 0) - 1 . \quad (32)$$

We use the two-particle probability density Eq. (24) and assume that not only the momenta of the test-particles are equal, i. e. $\mathbf{p}_1 = \mathbf{p}_2$, but that in addition also the wavepacket momenta are equal, i. e. $\mathbf{P}_1 = \mathbf{P}_2$. As a consequence the norm $\mathcal{N}_{12}$ Eq. (27) becomes independent of the momentum and the integration can be carried out easily. We obtain

$$\lambda = 1 + 2 \sum_{j=1}^{\infty} \frac{(-1)^j}{\left( 1 + \frac{jR^2}{2\sigma_o^2} \right)^{3/2}} . \quad (33)$$

It is not surprising that $\lambda$, and hence the number of configuration space cells $K$, is a function of the ratio $\frac{R^2}{2\sigma_o^2}$, i. e. the source dimension devided by the single-state dimension. For smaller wavepackets, the number of cells increases, the source becomes more incoherent and the chaoticity parameter $\lambda$ approaches 1.

The approximation $\mathbf{P}_1 = \mathbf{P}_2$ is responsible for the fact that $\lambda$ does not depend on the momentum coordinates. A more accurate treatment would yield that $\lambda$ is not independent on the momentum space distribution, even in case of static sources. We will demonstrate in Sect. 3 that Eq. (33) nevertheless provides a reasonable estimate for the chaoticity.

### 2.3 Physical legitimation of the generalized model

The wavepacket model seems to introduce a new parameter $\sigma_o$ into the physics of pion interferometry. Such a step has to be justified by physical arguments, i. e. it has to be demonstrated that the expansion of the formalism not only increases the complexity of the formula but also allows to describe additional observable phenomena which otherwise remain hidden.

In order to demonstrate the conceptual difference between the usual picture which leads to the standard correlation function Eq. (6) and the wavepacket model, we write, using the Fourier expansion Eq. (14), the two-particle state Eq. (25) in the form

$$|\Phi(\mathbf{p}_1, \mathbf{p}_2)|^2 = h^{-3} \times$$
$$\left| \int \int d^3x_1 \, d^3x_2 \, \left( \psi_1(\mathbf{x}_1) \, e^{-i\,\mathbf{p}_1 \cdot \mathbf{x}_1/\hbar} \, \psi_2(\mathbf{x}_2) \, e^{-i\,\mathbf{p}_2 \cdot \mathbf{x}_2/\hbar} \right. \right.$$
$$\left. \left. + \, \psi_2(\mathbf{x}_1) \, e^{-i\,\mathbf{p}_1 \cdot \mathbf{x}_1/\hbar} \, \psi_1(\mathbf{x}_2) \, e^{-i\,\mathbf{p}_2 \cdot \mathbf{x}_2/\hbar} \right) \right|^2 . \qquad (34)$$

In the next step, we replace the configuration space states by delta distributions:

$$\psi_i(\mathbf{x}_j) \longrightarrow \delta(\mathbf{X}_i - \mathbf{x}_j) , \qquad (35)$$

which yields

$$|\Phi(\mathbf{p}_1, \mathbf{p}_2)|^2 = h^{-3} \times$$
$$\left| e^{-i\,\mathbf{p}_1 \cdot \mathbf{X}_1/\hbar} \, e^{-i\,\mathbf{p}_2 \cdot \mathbf{X}_2/\hbar} + e^{-i\,\mathbf{p}_1 \cdot \mathbf{X}_2/\hbar} \, e^{-i\,\mathbf{p}_2 \cdot \mathbf{X}_1/\hbar} \right|^2 . \quad (36)$$

Now we have reached the plane wave picture in momentum space Eq. (1). After substituting this expression into Eq. (24) it is immediately obvious that the resulting expression is identical with the (incoherent) nominator of the correlation function [10], leading to Eq. (5). The substitution (35) therefore orthogonalizes the states in configuration space and also decouples them from momentum space, since $\psi_i(\mathbf{x}_j)$ contains momenta which are lost in $\delta(\mathbf{X}_i - \mathbf{x}_j)$. This procedure introduces a chaotic source (chaoticity $\lambda = 1$). On the other hand, in the same approximation the single-particle momentum distribution Eq. (20) would become the unity distribution. This is, as well as the obtained total chaoticity, in disagreement with the experimental results. To overcome these difficulties, the chaoticity $\lambda$ is introduced as a free parameter to account for partial coherent sources as a consequence of nonorthogonal states in configuration space. This at least in principle mimics the necessity to localize the particles in momentum space. In the usual treatment of pion correlations, however, the picture of localized particles is adopted and simultaneously the pions are forced to satisfy a certain energy distribution, which is a violation of Heisenberg's uncertainty principle and must be regarded as a semiclassical approximation of the physical reality.

The wavepacket picture, on the other hand, accounts for all quantum mechanical properties of the system in a quite transparent manner: The localization of the particles in configuration space leads to a delocalization in momentum space and vice versa. The chaoticity $\lambda$ is not a free parameter, but a function of other system parameters, especially of the geometrical dimensions in configuration space Eq. (33). From this point of view, $\sigma_o$ does not represent an additional new parameter, but instead a parametrization of $\lambda$.

It is an interesting question whether any physical meaning can be attributed to $\sigma_o$. In the opinion of the authors the concept of wavepackets is a natural consequence of the processes that occur in the fireball: A pion, which is created in a certain area of the fireball, has a small probability to migrate over large distances in the fireball. More specifically: Since there exists a mean free path $l$ for the pion to travel inside the fireball, it is effectively localized within a certain volume at freeze-out. The mean free path $l$ depends on the momentum, for 200 MeV/c pions and in nuclear matter with density $\rho_o$ it is expected to be $l = 1.3$ fm [18]. On the other hand, the density of the fireball at maximum compression is around $3\rho_o$ and near freeze-out it is found to be $1/3 \, \rho_o$ [19]. Hence the mean free path $l$ for 200 MeV/c pions in the fireball can be expected to be anywhere between 0.4 fm and 4 fm. Assuming that two pions interfere over a distance $\approx l$ and move independently in case of much larger distances, $l$ may roughly set the size of the localization volume and therefore of the wavepackets. Therefore, the measurement of the chaoticity $\lambda$ in connection with the source radius may, at least in principle, provide some information about the mean free path and hence the cross section for pion-absorption in nuclear matter.

## 3 Monte Carlo method

The procedure that will be presented in Sect. 3.1 is a generalization of Zajc's method [9] applied to the model of Sect. 2. The $n$-particle state is described in momentum space. The high complexity of the permanent Eq. (18) requires methods which are different from Zajc's approximation, since the latter one is only applicable in case of real valued wave functions. Instead we exploit the fact that as a consequence of the wave packet approach the particles are localized within a certain momentum space volume. Consequently, the wavefunction decays into clusters which do not substantially overlap, and therefore the symmetrization has to be carried out only clusterwise. This is described in detail in Sect. 3.2.

In Sect. 3.3 we demonstrate that clustering alone is not sufficient to reduce the numerical complexity in case of high momentum space densities. By means of simulations we investigate the impact of further factorizations on the correlation data.

### 3.1 Metropolis procedure

The method to generate an ensemble of test-particles with the statistical properties described in Sect. 2.2 is the standard Metropolis procedure [20]. At the beginning of each event only the phase space distribution Eq. (7) of wavepacket centres is defined. We now build a set of $n$ test-particles in momentum space. Although their initial momenta are arbitrary, some specific choices may accelerate the speed of convergence to the correct statistics. We choose the initial momenta in a way which is dictated by the single-particle wave functions, i. e. the coordinates are sampled within the Gaussian probability distributions defined by the packets $\phi_i(\mathbf{p})$ Eq. (15). The advantage is that the ensemble of particles immediately satisfies the single-particle momentum spectrum Eq. (20). There are still, of course, no correlations created. These are introduced by the following procedure:

- Evaluate the value of the $n$-particle density function $|\Phi(\mathbf{p}_1, \ldots, \mathbf{p}_n)|^2$
- Do ...
  - Change coordinate of one test-particle $\mathbf{p}_i \to \mathbf{p}_i + \Delta\mathbf{p}$
  - Evaluate density $|\Phi|^2_{\text{new}}$ with the new set of coordinates
  - Accept new configuration with probability $\mathcal{P} = \min\{1, |\Phi|^2_{\text{new}}/|\Phi|^2\}$
- ... Loop over all particles

After each particle was treated once, one *sweep* is finished. It needs several sweeps until the coordinates of test-particles have converged to values which are compatible with the momentum space probability distribution $|\Phi|^2$. How many sweeps are needed depends on the size $n$ of the system as well as on the specific properties of the density function. We found by "trial and error" that the rule

$$N_{\text{sweep}} = 30 \times n \tag{37}$$

yields results that do not change by further increasing the value of $N_{\text{sweep}}$. A second parameter of the simulation is the size of displacement $\Delta\mathbf{p}$. It is a well known rule [21] that these displacements must not exceed the natural fluctuations inherent in the system. Otherwise unphysical states have a large probability to be reached and the acceptance rate of the Metropolis selection becomes too small. The fluctuations are given by the widths $\sigma_p$ of the wavepackets in momentum space, and therefore we sample $\Delta\mathbf{p}$ within a uniform sphere of radius $\sigma_p$.

A series of simulations with different wavepacket widths $\sigma_o$ was carried out. Each simulation contained 150000 events of 5 pions. The phase space distribution of the wavepackets was chosen by the simple parametrizations Eq. (4) and Eq. (19), with temperature $T = 50$ MeV and source radius $R = 3$ fm. The results are displayed in Fig. 2.

The effective temperatures are in qualitative agreement with the calculation of the single-particle spectrum Eq. (20), but systematically smaller. Such a reduction is not observed in the single-pion system. It therefore seems that the interference phenomena present in the 5-pion system tend to decrease the pion energies relative to the expected single-particle spectrum. The effect is small ($\approx 4\%$ for $\sigma_o = 1.0$ fm) but significant and becomes more pronounced the smaller the wavepacket width is in configuration space, i. e. the larger the overlap in momentum space.

The correlation function is obtained by an eventwise evaluation of all pairs $\mathbf{p}_i - \mathbf{p}_j$. They contribute to the nominator of the correlation function. The denominator, that represents the uncorrelated 2-particle distribution, is obtained using event-mixing techniques. The standard correlation function Eq. (6) is fitted to the resulting distribution, yielding an effective radius $R_{\text{eff}}$ and a chaoticity $\lambda$. For comparison, the correlation function Eq. (2) was solved numerically using the states Eq. (24) and Eq. (20) for a large number of wavepacket widths. Again, the standard correlation function was used to obtain radius and chaoticity parameter for each solution, yielding the solid curve in the middle panel of Fig. 2. It is worth noting that the exact correlation function is not Gaussian, although the source density was chosen to be Gaussian. Figure 3 displays the correlation function corresponding to $\sigma_o = 1.8$ fm (circles) and the fit of the standard correlation function (solid curve). Clearly, the Gaussian fit-function is not the best choice, but in order to keep the results comparable with the experimental data we decided to avoid any modifications of the common analysis procedures. The difficulties in the fit procedures generate systematical errors, that in our simulations are larger than the statistical errors. In the figures, however, only the statistical errors are displayed.

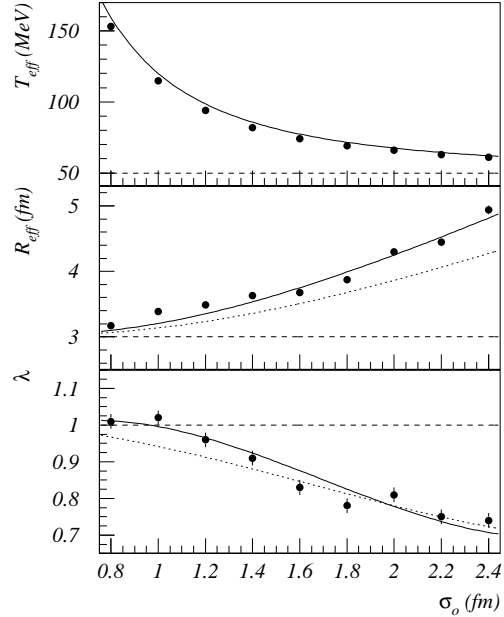

**Fig. 2.** The dependence of effective temperature $T_{\text{eff}}$ (upper panel), effective radius $R_{\text{eff}}$ (middle panel) and chaoticity $\lambda$ (lower panel) on the wavepacket widths $\sigma_o$. The circles represent the simulation results. Upper panel: The solid curve represents Eq. (21), the dashed line denotes the input temperature. Middle panel: The solid curve is in accordance to the exact solution of Eq. (24), the dotted curve is the approximation Eq. (29), the dashed line denotes the source radius. Lower panel: The solid curve is in accordance to the exact solution of Eq. (24), the dotted curve is the approximation Eq. (33) and the dashed line denotes the constant chaoticity obtained in the approximation Eq. (28).

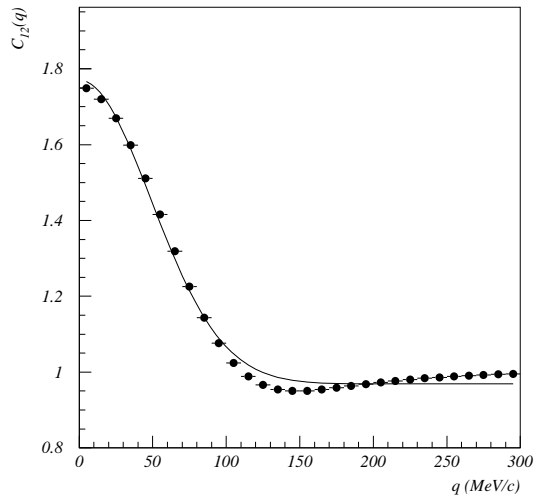

**Fig. 3.** The correlation function obtained by numerical (Monte Carlo) integration for the system used in Sect. 3 and $\sigma_o = 1.8$ fm (circles). The solid curve is the fit of the Gaussian standard correlation function Eq. (6).

The obtained radii are larger than the one expected using the zero'th order approximation (Eq. (29), dotted curve in the middle panel of Fig. 2), but in agreement with the exact solution of the correlation function (solid curve). It is interesting that the exact results of the effective radii can be parametrized with a function of the same type as Eq. (29), namely

$$R_{eff} = \sqrt{R^2 + \frac{\kappa\,\sigma_o^2\,T}{T_{eff}}}\,,\tag{38}$$

yielding $\kappa = 3.05 \pm 0.01$.

In accordance to the discussions presented in Sect. 2.2.2, the chaoticity $\lambda$ is not constant but decreases with increasing wavepacket width (lower panel). The solid curve represents the exact numerical solution, the dotted curve is the approximation Eq. (33) obtained by decoupling momentum- and configuration space. The accuracy of the approximation is expected to decrease in case of a strong coupling of momentum- and configuration space, i.e. flow. Quantitative estimates have to be obtained by extending the simulations.

The simulations were performed for a small system of only 5 pions. The generation of 150000 events required about 3 hours of CPU-time on a DEC $250^{4/266}$ workstation. For larger systems the CPU-time increases dramatically. One way to overcome these problems is to split the system up into smaller subsystems. This is described in the following section.

### 3.2 Cluster identification

The direct evaluation of the $n$-particle permanent Eq. (18) requires to sum up $n!$ terms. There are, of course, more intelligent algorithms available [22], but they still exhibit an exponential dependence of the complexity on the system size. Whereas in the case of determinants so called *elimination algorithms* like the *Gaussian elimination method* with a complexity $\sim n^3$ are available, a corresponding procedure for permanents seems to be out of reach [23]. Faced with the problem to evaluate systems of much more than 10 pions, we have to find a way to circumvent the necessity of computing the complete $n$-particle permanent. The solution can be found in terms of an *effective factorization*.

In order to understand this phenomenon, we consider the example of two particles. The probability density in momentum space is given by Eq. (25). Obviously the state factorizes if the second term $|\phi_2(\mathbf{p}_1)|^2 |\phi_1(\mathbf{p}_2)|^2$ and the interference term $\mathcal{I}$ vanish. Inserting Eq. (15) in Eq. (26) yields

$$Re\{\mathcal{I}\} = \exp\left(-\frac{1}{4\sigma_p^2}\left((\mathbf{P}_1 - \mathbf{p}_1)^2 + (\mathbf{P}_2 - \mathbf{p}_2)^2\right.\right.$$
$$\left.\left. + (\mathbf{P}_2 - \mathbf{p}_1)^2 + (\mathbf{P}_1 - \mathbf{p}_2)^2\right)\right)$$
$$\times \cos\left((\mathbf{X}_1 - \mathbf{X}_2)\cdot(\mathbf{p}_2 - \mathbf{p}_1)/\hbar\right)\,.\tag{39}$$

If the distance of the wavepackets $|\mathbf{P}_2 - \mathbf{P}_1|$ is much larger than the width $\sigma_p$, then necessarily one of the cross terms $(\mathbf{P}_i - \mathbf{p}_j)^2$ aquires a value much larger than $1/4\sigma_p^2$ so that the exponential term and consequently the interference vanishes. The same argument holds for the term $|\phi_2(\mathbf{p}_1)|^2 |\phi_1(\mathbf{p}_2)|^2$. Thus the states are separated and a symmetrization does not influence the probability density Eq. (25).

We want to exploit this result and split a given momentum space wavefunction into smaller parts. This means that clusters of overlapping wavepackets have to be identified. We can take advantage of the fact that the wavefunction in momentum space is stationary, c.f. Eq. (15), which means that this procedure has to be applied only once per event. First of all one has to define under which conditions two wavepackets factorize. We have found that reasonable results are obtained using a cut-off value of

$$|\mathbf{P}_2 - \mathbf{P}_1| < 2.5\,\sigma_p\,,\tag{40}$$

which implies that wavepackets with distances smaller than $2.5\,\sigma_p$ are defined to overlap, otherwise they factorize. The second step is to construct a matrix (*tableau*) which indicates whether two states overlap or not. Finally a tracer is launched to find a way through the tableau in order to locate the overlapping states.

If the momentum space density of the system increases, it may happen that the wavepackets form a very large cluster (*percolation*). Taking into account the fact that the symmetrization of clusters with sizes larger than $\approx 10$ takes too much time, the cluster has to be split up into subclusters by brute force, i.e. the wavefunction is forced to factorize. The influence of neglecting some of the interference terms on the correlation data is investigated in the following section.

### 3.3 Unauthorized factorization of wavefunctions

In order to keep the CPU-time within a tolerable range, it is necessary to factorize the wavefunctions of systems with very high momentum space densities, without taking care whether the criterion Eq. (40) is satisfied or not. To examine the influence of such an operation on the correlation function, a system of 30 pions was simulated under different conditions: The maximum size $N_{max}$ of the clusters was varied between 2 and 10. This means: If the cluster detection algorithm has identified a cluster of multiplicity $n$, then, if $n > N_{max}$, the cluster is divided into smaller subclusters. The effect of such an unauthorized factorization is of interest since most of the available codes generate pion correlations only in terms of 2-particle correlations Eq. (3). It is obvious that these methods become inaccurate if a certain phase space density is exceeded and correlations of higher order than 2 begin to dominate.

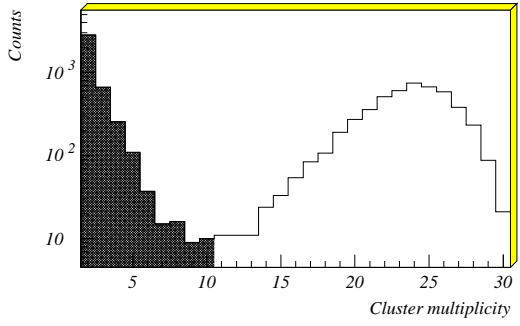

**Fig. 4.** The distribution of cluster multiplicities identified by the cluster-tracer. Only a subset of them (here: clusters of size $\leq 10$, shadowed) can be treated in full symmetry, the rest has to be split up into smaller parts.

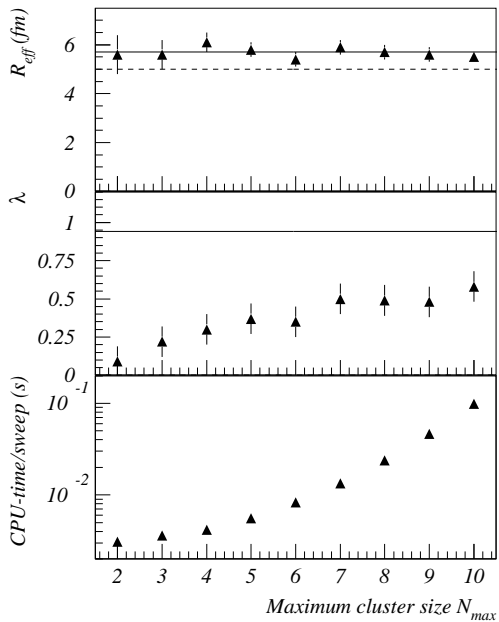

**Fig. 5.** Simulation results (triangles) for different maximum cluster sizes $N_{max}$. Upper panel: The effective correlation radii. The solid line denotes the result of the numerical evaluation of Eq. (24), the dashed line is the source radius. Middle panel: The chaoticity $\lambda$. Again, the solid line denotes the result of Eq. (24). Lower panel: The CPU-time per sweep.

In our simulation, the source was parametrized using the standard Gaussian Eq. (4) with radius $R = 5$ fm, the momentum distribution was the relativistic Maxwellian

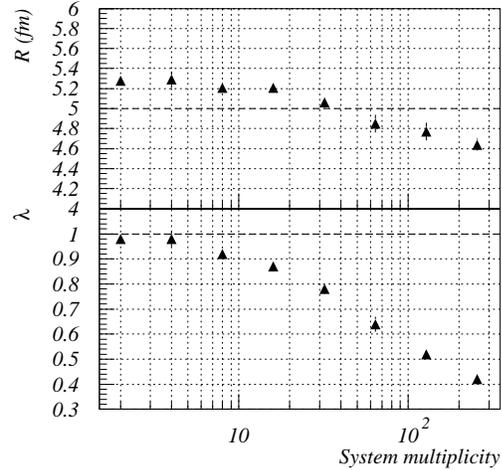

**Fig. 6.** Simulation results (triangles) for different system multiplicities. Upper panel: The correlation radii. The dashed line denotes the input radius. Lower panel: The chaoticity $\lambda$. The dashed line denotes the expected value $\lambda = 1$, corresponding to the chaotic source.

$$\mathcal{P}(\mathbf{P}) \sim \exp\left(-\frac{\sqrt{\mathbf{P}^2 + m^2}}{T}\right) \qquad (41)$$

with temperature $T = 50$ MeV. The wavepacket widths were $\sigma_o = 1.8$ fm. Figure 4 shows the distribution of cluster multiplicities. A large amount of clusters with sizes of around 25 particles was found, indicating that percolation phenomena play a significant role. Figure 5 displays the effects appearing if the clusters are broken into pieces of $N_{max} = 2, 3, \ldots, 10$. The upper panel shows the effective radii obtained in the same way as described in Sect. 3.1. The forced factorization obviously has no influence on the extracted radii. These also agree with the numerical solution of the 2-particle correlation function Eq. (24), which yields $R_{eff} = 5.70 \pm 0.05$ fm (solid line). In the middle panel a strong dependence of the chaoticity parameter $\lambda$ on $N_{max}$ is displayed (triangles). The result of the numerical solution of Eq. (24) ($\lambda = 0.94 \pm 0.01$, solid line) is not reached by any simulation. This implies that the neglection of interference terms in the simulation procedure makes an accurate determination of the chaoticity (and therefore of the number of degrees of freedom in configuration space) impossible. This has to be taken into account when only 2-particle correlations are used to *simulate* events.

There exists an additional effect which is described by Zajc [9]: In case of high phase space densities also the *analysis* suffers from the fact that the 2-particle correlation function ignores the contributions of higher order correlations. This phenomenon leads to a reduced value for $\lambda$ and the effective radius. In order to study the influence quantitatively, a series of simulations with increasing system size was carried out using the quantum statistical code presented in [8]. This code is based on the plane wave approximation (35), i. e. it *per definition* generates pions from a chaotic source and yields the correlation radius $R_{eff} = R$. On the other hand, it includes Bose-symmetrization up to all orders. Thus, this method offers the best choice to study the effects which appear exclusively in the analysis procedures. In the simulations, the same phase space parametrization as above ($R = 5$ fm, $T = 50$ MeV) was used, the results are presented in Fig. 6. The fact that the radius comes out somewhat too large for all simulations is a consequence of the discretization of the momentum space into cells. In agreement with the predictions of Zajc, the radius as well as $\lambda$ decrease with increasing system size, for the latter the effect is more pronounced. For a system of 30 pions this is already visible and therefore the extracted $\lambda$ values in Fig. 5 are expected to saturate at approximately $\lambda(N_{max} = 30) \approx 0.75$ instead of 0.94. We want to emphasize that here, in contrast to the simulations shown in Fig. 6, the deviation of $\lambda$ from 1 is due to a combination of three effects, (1) the omission of interference terms in the simulation, (2) the omission of higher order correlations in the analysis and (3) partial coherence, where the contributions from (2) and (3) are independent on $N_{max}$. The contribution from (2) also depends on the phase space volume, because the number of momentum space cells is proportional to the product $R^2 T$ [8]. The larger the number of cells, the smaller the probability to find more than two pions in identical cells, which implies that the amount of reduction becomes smaller. For a system similar to the 200 AGeV Pb + Pb experiment in CERN (e. g. $R = 6$ fm, $T = 150$ MeV, pion multiplicity: 1000), we obtain $R = (5.6 \pm 0.1)$ fm and $\lambda = 0.75 \pm 0.04$ for a chaotic source. This implies that for an accurate measurement of the chaoticity higher order correlation functions have to be included.

Finally, the lower panel in Fig. 5 displays the needed CPU-time per sweep. It is clear that the evaluation of clusters with $N_{max}$ much larger than 10 is prohibited by the exponential growth of the operations to be carried out by the algorithm. The generation of 2000 events under $N_{max} = 10$ - conditions needed about 15 hours on a DEC $250^{4/266}$ workstation.

In this and the foregoing section two ways of reducing the calculational effort in the simulation were presented:

1. Truncation of the $n$-particle wavefunction into clusters in accordance with the criterion of effective factorization. This operation does not effect the correlation properties.
2. A split up of the clusters into subclusters of sizes $\leq N_{max}$. This operation is unauthorized in the sense that interference terms are lost. Although the correct source radii can be extracted even in case of $N_{max} = 2$, the chaoticity $\lambda$ in the correlation function decreases dramatically.

In addition, the factorization of the system by *analyzing* only 2-particle correlations also effects $\lambda$ in a way that an accurate measurement of the chaoticity under e. g. CERN conditions becomes possible only when higher order correlation functions are used.

## 4 Molecular dynamic method

The Monte Carlo procedure presented above allows to study a large class of physical scenarios, including collective flow and dynamic changes of the source. However, these situations are correctly treated only when the pions can be regarded as free, which implies that there are no forces acting on them after freeze-out. This is due to the fact that the Monte Carlo procedure is not capable of providing a physical timescale, instead it is designed to generate statistical ensembles under the condition of stationarity.

On the other hand, final state interactions of long range can occur in heavy ion collisions, for example

1. the Coulomb interaction between pions and the central fireball,
2. the Coulomb interaction between pion and pion,
3. the Coulomb interaction between pions and the spectators.

These interactions require a time dependent treatment in configuration space, i. e. an integration of the equation of motion. This is the domain of molecular dynamic procedures.

At this point we are confronted with a fundamental problem: *Quantum "mechanics" does not provide us with an equation of motion.* Instead it contains a number of rules for the time development of probability densities. Thus the evolution of a system has to be found by a numerical solution of the $n$-particle Schrödinger equation, which is very time consuming and inadequate to our goal. It is remarkable that there exists an extension of the usual quantum mechanical formalism [24] (for a thorough treatment including applications see [26]) that allows us to formulate rules for the motion of certain test-particles quite similar to the particles used in the Monte Carlo method. In the following section the application of this formalism to our free-particle model (Sect. 2) is demonstrated, Sect. 4.2 contains an approximative treatment of Coulomb interaction between pions and fireball.

### 4.1 Bohm's quantum theory of motion

The system in configuration space is described by the $n$-particle wave function Eq. (13). We consider a set of $n$ test-particles with phase space coordinates

$$\{(\mathbf{x}_1(t), \mathbf{p}_1(t)), (\mathbf{x}_2(t), \mathbf{p}_2(t)), \dots, (\mathbf{x}_n(t), \mathbf{p}_n(t))\} \ . \quad (42)$$

Further we may choose a set of initial conditions for the configuration space coordinates of the test-particles

$$\mathcal{P}(\mathbf{x}_1, \mathbf{x}_2, \dots, \mathbf{x}_n, 0) = |\Psi(\mathbf{x}_1, \mathbf{x}_2, \dots, \mathbf{x}_n, 0)|^2 \quad (43)$$

as well as for the momentum space coordinates

$$\mathbf{p}_j(0) = \nabla_j S(\mathbf{x}_1, \dots, \mathbf{x}_n, 0) \ , \quad (44)$$

where

$$S = \frac{\hbar}{2i} \log\left(\frac{\Psi}{\Psi^*}\right) \quad (45)$$

is the phase of the wavefunction and $j = 1, 2, \ldots, n$. If the wavepackets are small enough so that their overlap can be neglected, the initial distribution reduces to

$$\mathcal{P}(\mathbf{x}_j, 0) = |\psi_j(\mathbf{x}_j, 0)|^2 \tag{46}$$

and

$$\mathbf{p}_j(0) = \mathbf{P}_j . \tag{47}$$

It was shown by Bohm [25] that application of the following equation of motion to the test particles yields in the limit $t \to \infty$ a phase space distribution which coincides with the configuration space *and* momentum space density Eq. (18):

$$m \frac{d^2 \mathbf{x}_j(t)}{dt^2} = -\nabla_j Q(\mathbf{x}_1, \mathbf{x}_2, \ldots, \mathbf{x}_n, t) \tag{48}$$

with

$$Q = \sum_{k=1}^{n} -\frac{\hbar^2}{2m} \frac{\nabla_k^2 R}{R} . \tag{49}$$

$R = (\Psi^* \Psi)^{1/2}$ denotes the amplitude of the wavefunction. Obviously, Eq. (48) is an equation of motion of Newton type, but the potential $Q$, called *quantum potential*, is nonlocal since it contains the (symmetrized) amplitude. We want to emphasize that

1. the equation of motion Eq. (48) is not based on a semiclassical approximation but is of pure quantum mechanical nature,
2. Bohm's picture of quantum mechanics leads to the same observable predictions for the test-particles as the conventional approach for the results of measurements. Thus all ensemble properties obtained in Sect. 2.2 remain valid for the test-particles.

The quantum potential Eq. (49) for the wavepacket model can be expressed analytically but exhibits a quite complicated structure for many-particle systems. Here we only want to discuss the most simple case of one particle. Considering a wavepacket Eq. (11) at rest ($\mathbf{P} = 0$) and centred at $\mathbf{X} = 0$ we obtain

$$Q(\mathbf{x}, t) = \frac{\hbar^2}{4m\sigma_o^2} \left( 3 - \frac{\mathbf{x}^2(t)}{2\sigma^2(t)} \right) , \tag{50}$$

where $\sigma(t) = \sqrt{s(t) s^*(t)}$ is the time dependent amplitude of the complex width $s(t)$ Eq. (12). If we are interested in the average initial quantum potential of a large ensemble of single-particle systems, we have to integrate over the wavepacket-density distribution

$$|\psi(\mathbf{x}, t=0)|^2 = (2\pi\sigma_o)^{-3/2} \exp\left( \frac{-\mathbf{x}^2}{2\sigma_o^2} \right) , \tag{51}$$

obtaining

$$\langle Q(t=0)\rangle = \int Q(\mathbf{x}, 0) \, |\psi(\mathbf{x}, 0)|^2 \, d^3x = \frac{3\hbar^2}{8m\sigma_o^2} , \tag{52}$$

and comparing with Eq. (21) one finds

$$\langle Q \rangle = \frac{3}{2} T_q . \tag{53}$$

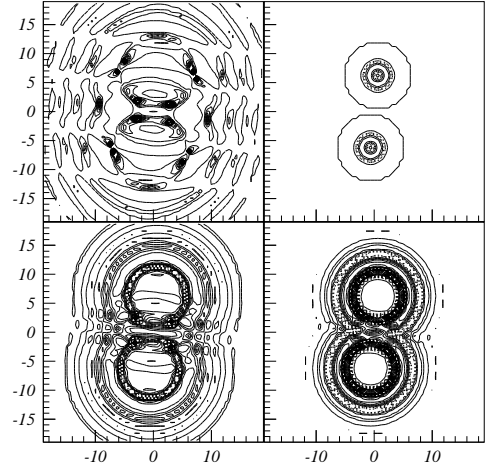

**Fig. 7.** Contour plots of the quantum potential corresponding to a 3-particle system. Beginning with the upper right panel and proceeding clockwise, the pictures display the potential after 0, 10, 15 and 25 fm/c. Two wavepackets were fixed in the $xy$-plane, a third was used to scan the plane. The labels are in fm-units.

This implies that the quantum potential in Bohm's picture of individual particles manifests itself (in the ensemble picture) in terms of the quantum temperature we have found to be a consequence of Heisenberg's uncertainty in momentum space. How this determines the dynamics of the particles can be seen by looking at the quantum force

$$-\nabla Q = \frac{\hbar^2}{4m\sigma^4(t)} \mathbf{x}(t) . \tag{54}$$

If the particle is placed at the centre of the wavepacket ($\mathbf{x}(0) = 0$), the quantum force disappears and the particle follows a classical trajectory (Ehrenfest's theorem). Otherwise, the force always points away from the wave packet centre, i. e. in a large ensemble the test-particles are accelerated isotropically in all directions, indicating the dispersion of the wave packet. In addition, the force decreases rapidly with increasing spreading ($\sigma^{-4}$ dependence). What we can learn from this example is that the quantum force accounts for Heisenberg's uncertainty principle, but from an individual rather than from an ensemble average point of view.

To get an impression, Fig. 7 shows the quantum potential of a 3-particle system and it's time development. To obtain the pictures, two wavepackets with initial widths $\sigma_o = 2.0$ fm and centred test-particles were fixed in the $xy$-plane. Since they are at rest, the only dynamics visible is due to dispersion. A third wavepacket (with centred test-particle) is used to scan the $xy$-plane at different times. Therefore, the pictures display the potential seen by the test-particle of the third wavepacket.

For numerical simulations, there exists a different and less expensive way to solve the equation of motion: The velocities of the particles can be obtained by Eq. (44), which we can rewrite as

$$\mathbf{v}_j(t) = \frac{\hbar}{m\,|\Psi|^2}\, Im\left\{\Psi^*\nabla_j\Psi\right\}\;. \qquad (55)$$

The force acting on the particle in the discrete time interval $dt = t_n - t_{n-1}$ can be expressed as $-\nabla Q = m\,d\mathbf{v}/dt$ with $d\mathbf{v} = \mathbf{v}(t_n) - \mathbf{v}(t_{n-1})$. The gradient of the permanent $\Psi$ is obtained in the following way:

$$\begin{aligned}
\nabla_k\Psi &= \nabla_k\left(\sum_\sigma \psi_{\sigma(1)}(\mathbf{x}_1)\,\psi_{\sigma(2)}(\mathbf{x}_2)\cdots\psi_{\sigma(n)}(\mathbf{x}_n)\right)\\
&= \sum_\sigma \nabla_k(\psi_{\sigma(k)}(\mathbf{x}_k))\,\psi_{\sigma(1)}(\mathbf{x}_1)\cdots\psi_{\sigma(k-1)}(\mathbf{x}_{k-1})\\
&\qquad \times\psi_{\sigma(k+1)}(\mathbf{x}_{k+1})\cdots\psi_{\sigma(n)}(\mathbf{x}_n)\\
&= \sum_{i=1}^n \nabla_k(\psi_i(\mathbf{x}_k))\left(\sum_{\sigma\neq i}\psi_{\sigma(1)}(\mathbf{x}_1)\cdots\psi_{\sigma(k-1)}(\mathbf{x}_{k-1})\right.\\
&\qquad \left.\times\psi_{\sigma(k+1)}(\mathbf{x}_{k+1})\cdots\psi_{\sigma(n)}(\mathbf{x}_n)\right)\\
&\equiv \sum_{i=1}^n \nabla_k(\psi_i(\mathbf{x}_k))\,\Psi^{(i,k)}(\mathbf{x}_1,\ldots,\mathbf{x}_n)\;, \qquad (56)
\end{aligned}$$

where $\Psi^{(i,k)}(\mathbf{x}_1,\ldots,\mathbf{x}_n)$ is the permanent of the matrix after canceling the $i$'th row and $k$'th column in $\Psi(\mathbf{x}_1,\ldots,\mathbf{x}_n)$. We note that gradients of permanents are sums of permanents of the submatrices. For the computation of the permanents the code presented in [22] is used.

It is clear that the presence of such irregular potentials as displayed in Fig. 7 requires very small timesteps for the integration of the equation of motion. This simulation method therefore proved to be usable only for small systems of less than $\approx 10$ particles. However, there is some kind of separability inherent in the system: Figure 7 suggests that the quantum potentials of the single wavepackets disturb each other only in case of overlap, and since the packets move away from each other in a realistic simulation, it is possible to divide the wavefunction into subparts in the same way as described in Sect. 3.2. The separation procedure has to be repeated frequently, since the configuration changes with the progress of time.

### 4.2 Simulation including Coulomb interaction

It was our goal to include final state interactions into the simulation, and we present the example of the pion–fireball Coulomb interaction. For this purpose we describe the charge distribution by a homogeneous sphere of rms-radius $r = 8.6$ fm and total charge $Z = 100$, corresponding to 100 positive nucleons in the participant region. For the equation of motion, we use the following approximation:

$$m\frac{d^2\mathbf{x}_j}{dt^2} = -\nabla_j(Q+V)\;, \qquad (57)$$

where $V$ denotes the classical (static) Coulomb potential. The wavepackets are also exposed to the Coulomb field, because they represent a cloud of charge $Z = 1$, but their shapes remain uneffected. This picture is correct only as far as the Coulomb field is homogeneous on the scale of the wavepacket diameter. Otherwise the quantum potential becomes a function of $V$. In this sense, $Q$ in Eq. (57) can be regarded as the zero'th order Taylor approximation of $Q(V)$.

For the simulations, wavepackets with initial widths $\sigma_o = 1.8$ fm were used. Again, the phase space distribution of the wavepackets was obtained by the parametrizations Eq. (4) with radius $R = 7$ fm (in accordance to the rms-radius $r = 8.6$ fm of the charged sphere) and Eq. (41) with temperature $T = 50$ MeV. The initial phase space distributions of the test-particles were chosen in accordance to conditions Eq. (43) and Eq. (47), which is simple as long as the configuration space density is sufficiently small that the overlap of the initial wavepackets can be neglected. If this is not the case, a Monte Carlo procedure similar to the one discussed in Sect. 3 has to be applied to obtain the initial configuration space distribution Eq. (43).

In the simulations, $10^5$ events for neutral, positive and negative pions were generated. The adaptive time step was chosen in a way that the energy gain (resp. loss) of one particle did not exceed 1 MeV per timestep. Pion-pion interactions were not taken into account, but can be, of course, included in a similar way as the pion-fireball interaction. Each event contained a number of 6 pions, the CPU-time needed was about 1 s per event on a DEC 250$^{4/266}$ workstation.

While the quantum forces can aquire quite high values (up to several hundreds of MeV/fm) during the first phase of the evolution, they tend to decline rapidly in the course of the wavepacket dispersion (see Eq. (54) and Fig. 7), so that there exists a moment when they vanish and the equation of motion becomes classical. We found that a cut-off value of $|\mathbf{F}| = 0.01$ MeV is a sensible choice to stop the further evaluation of the event, since from that point on the momentum space distribution of the test-particles does not change substantially. For the neutral particles, an average number of 580 timesteps was needed to reach this point (which we may call "quantum freeze-out"), in terms of the system time this value corresponds to $\approx 130$ fm/c, but both values can fluctuate substantially from event to event. In case of charged pions, these values are higher due to the long range character of the Coulomb force.

Figure 8 displays the resulting normalized momentum spectra. It is not surprising that the spectra of the $\pi^+$ ($\pi^-$) are somewhat enhanced (suppressed) for low momenta in comparison to neutral pions, a consequence of the energy gain (loss) due to the central Coulomb potential. The results of the correlation analysis are displayed in Table 1. The values were obtained by fitting the standard correlation function Eq. (6) to the correlation signal Eq. (2) extracted from the simulated data. Within the errors, the $\pi^o$ - radii are in agreement with the results obtained by the numerical evaluation of Eq. (24), labeled as $\pi^o$(th.) in Table 1. The $\pi^-$ data, however, exhibit a significant radius increase for low momentum pions and approach the value of the nominal radius for high momenta. Integrated over all momentum slices, the $\pi^-$ - radius is about 20 % larger than the $\pi^o$ - radius. On the other hand, the $\pi^+$ - radii appear slightly enhanced in the low momentum range and drop below the $\pi^o$ values in the high momentum range. The integrated radius agrees, within the errors, with the integrated $\pi^o$ - radius. Within the statistical errors, the Coulomb interaction between charged pions and the fireball does not appear to effect the chaoticity parameter $\lambda$ significantly.

**Table 1.** Extracted radii $R_{eff}$ and chaoticity parameter $\lambda$ for neutral pions ($\pi^o$), positive charged ($\pi^+$) and negative charged pions ($\pi^-$) and for different momentum cuts. For comparison, results of the numerical evaluation of Eq. (24) are displayed in the last column.

| $p$ (MeV) | | $\pi^o$ | $\pi^+$ | $\pi^-$ | $\pi^o$ (th.) |
|---|---|---|---|---|---|
| 0 – 600 | $R_{eff}$ | 7.7±0.3 | 7.3±0.3 | 9.1±0.3 | 7.44±0.04 |
| | $\lambda$ | 1.0±0.1 | 1.0±0.1 | 0.9±0.1 | 0.96±0.01 |
| 0 – 120 | $R_{eff}$ | 7.3±0.5 | 8.1±0.4 | 10.8±0.8 | 7.35±0.04 |
| | $\lambda$ | 1.0±0.1 | 1.1±0.2 | 1.0±0.2 | 0.96±0.01 |
| 120 – 200 | $R_{eff}$ | 7.5±0.5 | 7.1±0.4 | 8.2±0.6 | 7.43±0.04 |
| | $\lambda$ | 0.9±0.1 | 0.9±0.1 | 0.8±0.1 | 0.96±0.01 |
| 200 – 600 | $R_{eff}$ | 7.0±0.6 | 6.4±0.5 | 7.5±1.0 | 7.44±0.04 |
| | $\lambda$ | 1.0±0.2 | 0.8±0.1 | 0.9±0.2 | 0.96±0.01 |

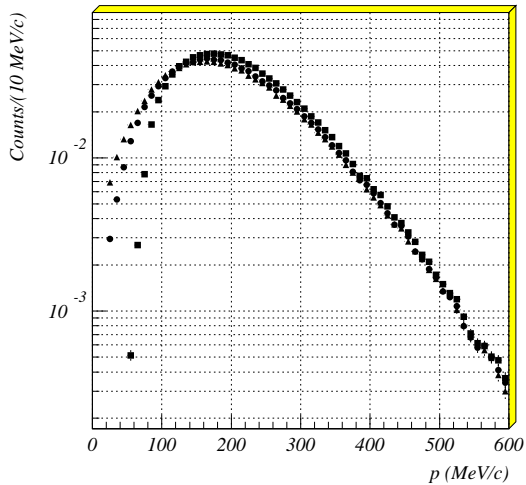

**Fig. 8.** Normalized momentum spectra for $\pi^o$ (circles), $\pi^+$ (squares) and $\pi^-$ (triangles).

It is an open question whether or not a more accurate treatment of the quantum potential, i. e. the inclusion of higher order Taylor terms in Eq. (57), may change the results presented in Table 1. But they may indicate the direction in which the deduced radii will change when the Coulomb interaction between the pions and the source is no longer negligible. They are compatible with the results obtained by Barz [27], who has solved the Klein-Gordon equation with Coulomb potential numerically by using a partial wave expansion technique. Bohm's the-ory provides an alternative way to approach the quantum mechanics of a system of particles with residual interactions. Approximative methods like perturbation theory can be applied to the quantum potential, which determines the dynamic changes of the system on a fundamental level.

## 5 Conclusion

In this work we have developed a method which transforms a given classical distribution of particles in phase space into a corresponding Bose-Einstein quantal distribution. In Sect. 2 we have demonstrated how this can be achieved by first building a symmetrical state from the original particle distribution and then introducing a set of test particles as a representative of the quantum mechanical ensemble. We have shown in Sect. 2.2 how the statistical properties of the resulting mixed ensemble deviate from the original classical ensemble: As a consequence of Heisenberg's uncertainty principle single particle distributions in configuration- and momentum space are smeared out. The relevant parameter is the initial wavepacket width $\sigma_o$, which represents the degree of information loss one is willing to accept either in configuration space or momentum space.

The two-particle correlation function for wavepackets is complicated and has to be treated numerically, but we have demonstrated that even for the simple example of a static, Gaussian and thermal source

1. the correlation function is not Gaussian (Eq. (24) and Fig. 3),
2. the correlation radius disagrees with the geometrical radius (Eq. (29)),
3. and the chaoticity $\lambda$ is a complicated function of the system parameters, in particular it depends on the extensions of the wavepackets relative to the source size (Eq. (33) and Fig. 2).

In Sect. 2.3 we have shown that the usual plane wave approach discouples the configuration space from the momentum space. It violates Heisenberg's uncertainty principle and does not provide a physical explanation for the chaoticity $\lambda$. In contrast, the wavepacket picture allows to relate the observable chaoticity with the localization of the pions in the source, i. e. their mean free path. This implies that $\lambda$ may contain information about the cross section for pion absorption in nuclear matter.

In Sect. 3 we presented a Monte Carlo procedure to treat systems of free (noninteracting) particles in momentum space. The consistency of the results with the expected ensemble properties was tested in Sect. 3.1. One interesting result is that the temperature deduced from the measured momentum spectrum of pions is smaller than the theoretical temperature of the single particle spectrum. A plausible reason would be the reduction of the degrees of freedom caused by the interference.

The large numerical effort needed to evaluate systems of large pion numbers motivated us to develop criteria under which the system may separate. In case of effective factorization, the correlation phenomena remain undisturbed (Sect. 3.2). However, the study of high momentum space densities requests a further, unauthorized factorization of the states. We have shown that also this factorization does not lead to a noticeable change of the effective correlation radii, but the chaoticity $\lambda$ becomes smaller as a consequence of omitting interference terms (Sect. 3.3). Therefore we argued that usual codes which only use pion pairs to generate correlated states are not capable of treating the chaoticity in a consistent way. We further want to point out that our approach to parametrize the system in terms of the standard pion source and the Maxwell distribution and to use the standard correlation function is an oversimplification, since it contains no more physics than "radius", "temperature" and "chaoticity". Phenomena like flow, time dependencies of the source, resonance decay and final state interactions should be included since they are readily observed in heavy ion collisions. The model we have presented is capable to contain these cases since it can be applied to any phase space distribution. An adequate analysis of these phenomena, however, may ask for finer observables than just 2-particle correlations. As an example we have demonstrated that an accurate measurement of the chaoticity $\lambda$ under conditions realized in the CERN Pb + Pb (200 AGeV) experiment, i. e. multiplicities in the order of 1000 pions, requires the analysis of higher order correlation functions.

Interactions of long range such as the Coulomb interaction can not be treated exclusively in momentum space. Instead, a time dependent calculation in configuration space in the framework of molecular dynamics has to be performed. The extension of the usual quantum mechanics given by Bohm offers a method to proceed from the laws of ensemble averages to the laws of a single representative (Sect. 4.1). The quantum forces, however, are of quite irregular nature and therefore a numerical treatment will usually be restricted to systems of low multiplicities.

The treatment of pion–fireball Coulomb interaction in Sect. 4.2 should be regarded as a first attempt to handle final state interactions without neglecting the quantum mechanical features of the system. The resulting discrepancy between the correlation radii of $\pi^+$ respectively $\pi^-$ pairs, which is largest for low pion momenta, has been observed in experiments [19]. Our simulations indicate that the effects on the radii are not of opposite size for the two pion charges, i. e. taking the average of both values as estimate for the source radius is not an adequate choice.

## 6 Appendix

Be $K$ the number of cells, $N$ the number of pions which are already distributed into the cells, $m$ the number of pairs, and $m_c$ the number of coherent pairs, i. e. the number of pairs that can be built from pions which occupy identical cells. We define $\mathcal{P}_c$ as probability to find a coherent pair, $n(i)$ the number of pions in the $i$'th cell and

$$w(i) = \frac{n(i) + 1}{K + N} \qquad (58)$$

the statistical weight of the $i$'th cell. We want to show that $\mathcal{P}_c = 2/(K + 1)$, independent on $N$.

Be $N = 1$. The occupied $j$'th cell has the statistical weight $w(j) = 2/(K+1)$, whereas all other cells have the weights $w(i) = 1/(K + 1)$, $i \neq j$. For the second pion, we therefore obtain

$$\mathcal{P}_c(N = 2) = w(j) = \frac{2}{K + 1} . \qquad (59)$$

We now assume that Eq. (59) holds for a certain $N \geq 2$, which implies that

$$\mathcal{P}_c(N) = \frac{m_c}{m} = \frac{\sum_{i=1}^{K} n(i)(n(i) - 1)/2}{N(N - 1)/2} \equiv \frac{2}{K + 1} . \qquad (60)$$

We now consider the $N + 1$'th pion and calculate how many coherent pairs it is expected to build with the other pions:

$$\langle m_c \rangle_{\mathrm{new}} = \sum_{i=1}^{K} w(i)\, n(i) = \sum_{i=1}^{K} \frac{(n(i) + 1)n(i)}{K + N} . \qquad (61)$$

We therefore obtain

$$\mathcal{P}_c(N + 1) = \frac{m_c + \langle m_c \rangle_{\mathrm{new}}}{m_{\mathrm{new}}}$$

$$= \frac{\sum_{i=1}^{K} \left( \frac{n(i)(n(i)-1)}{2} + \frac{(n(i)+1)n(i)}{K+N} \right)}{\frac{N(N-1)}{2} \frac{(N+1)}{(N-1)}}$$

$$= \frac{\sum_{i=1}^{K} \frac{(n(i)-1)n(i)}{2} \left( 1 + \frac{2}{K+N} \right) + \frac{2N}{K+N}}{\frac{N(N-1)}{2} \frac{(N+1)}{(N-1)}} \qquad (62)$$

where we have used that $\sum_{i=1}^{K} n(i) = N$. If we now substitute Eq. (60), we get

$$\mathcal{P}_c(N + 1) = \frac{2}{K + 1} \left( 1 + \frac{2}{K + N} \right) \frac{(N - 1)}{(N + 1)}$$

$$+ \frac{4}{(K + N)(N + 1)}$$

$$= \frac{2}{K + 1} . \qquad (63)$$

*Acknowledgement.* This work was supported by the Bundesministerium für Forschung und Technologie under contract No 06 HD 525 I and by the Gesellschaft für Schwerionenforschung mbH under contract No HD Pel K.